# Approaches for user profile Investigation in Orkut Social Network

Rajni Ranjan Singh, Deepak Singh Tomar

Maulana Azad National Institute of Technology (MANIT)

Bhopal, India.
ranjansingh06@gmail.com, deepaktomar@manit.ac.in

*Abstract-* **Internet becomes a large & rich repository of information about us as individually. Any thing form user profile information to friends links the user subscribes to are reflection of social interactions as user has in real worlds. Social networking has created new ways to communicate and share information. Social networking websites are being used regularly by millions of people, and it now seems that social networking will be an enduring part of everyday life. Social networks such as Orkut, Bebo, MySpace, Flickr, Facebook, Friendster and LinkedIn, have attracted millions of internet user who are involved in bogging, participatory book reviewing, personal networking and photo sharing. Social network services are increasingly being used in legal and criminal investigations. Information posted on sites such as Orkut and Facebook has been used by police, probation, and university officials to prosecute users of said sites. In some situations, content posted on web social network has been used in court. In the proposed work degree of closeness is identified by link weight approaches and information matrices are generated and matched on the basis of similarity in user profile information. The proposed technique is useful to investigate a user profile and calculate closeness /interaction between users.**

## I. INTRODUCTION

Orkut is one of the earlier & most famous web social networks run by google plays an important role to communicate and share private and public information in web environment facilitate bogging (scraping), personal networking, photo sharing, chatting, private messing, friend search. An interesting part of Orkut SNS is that user can see not only others profile information but also others friends networks. Recent work has attempted to find of web pages communities by performing analysis on their graph structure [1], Mining Directed Social Network from Message Board [2], Evaluating Similarity measure in Orkut networks [3]. Discover behaviour of Turkish people in Orkut[4], trust based recommendation system.our work focuses on individuals' homepages and the connections between them we can now use it to characterize relationships between people. Beyond developing the interface, we quantitatively evaluated the matchmaking approach for all kinds of information about the user. To predict whether one person is a friend of another & how much closeness both has, we rank all users by their similarity to that person. Intuitively, our matchmaking approach guesses that the more similar a person is, the more likely they are to be a friend. Similarity is measured by analysing profile information, mutual friends, and mutual communities. If we are trying to evaluate the likelihood that user A is linked to user B, we sum the number of items the two users have in common. Similarity between profiles reflects closeness and interaction between users .We only considered direct friends to the candidates for matching. Computing the similarity score for individual to all others friends in direct links, and rank the others according to their similarity score. We expect some friends to be more similar than others. The rest of the paper is organized as follows. In section 2 backgrounds in which we discuss about Orkut networks, user profiles, and friends networks In Section 3 we discuss approaches for user profile investigation Section 4 illustrate proposed framework for relation identification on the basis of profile similarity. Section 5 shows Experimental results, Section 6 discuss Challenges, Section 7 Conclusion and Section 8 shows References used in this paper.

## II. BACKGROUND

*A. Orkut an Overview*

Orkut is our topic of interest is a free-access social networking service owned and operated by Google. The service is designed to help users met new friends and maintain existing relationships It is one of the most visited websites in India and Brazil.

*B .Crime over Orkut*

Now Orkut social N/W are targeted by criminals and terrorist to spread wrong information plan blast recently many abduction terrorist activity are noticed by cyber police, now terrorist use the internet and tools like E-mail, Chat and social N/wing sites to plant terrorist attack.

*C. User profile: User profile is an individual user home page consists of mainly.*

(1) Social, professional, personal information.
(2) Links to other friends profile called friend list.
(3) Communities
(4) Photograph of user.
(5) Scrapbook
(6) Photo album

Fig. 2 Communities

TABLE 1

SITE FEATURES OF ORKUT SOCIAL NETWORKS

| Site Feature | Orkut |
|---|---|
| Profiles | Publicly viewable profile |
| Advertisement | Yes |
| Interface | Simply and easy to Understand |
| Chat | Yes, Google chat |
| Search | Yes, Google search |
| Customizable | No |
| Online/offline communication (Orkut scrapbook) | Support multiple language |
| Friends rating and profile view | Yes u can rate your friends |

*D. Friends Networks from User Profiles.*

The friend's network of Orkut, our topic of study, has two varieties of accounts: users and communities.

TABLE 2

USER, COMMUNITY, LINK RELATIONSHIP

| Start | End | Link Denotes |
|---|---|---|
| User | User | Trust or friendship |
| User | Community | Readership or Subscribership |
| Community | User | Membership, Posting access, maintainer |
| Community | Community | Obsolete |

*1). Friendlist:* - A user has connection with their friends by maintaining friendlist consist of links of all friends profiles reflects user social relation. User explicitly adds friends by accepting friend's request.

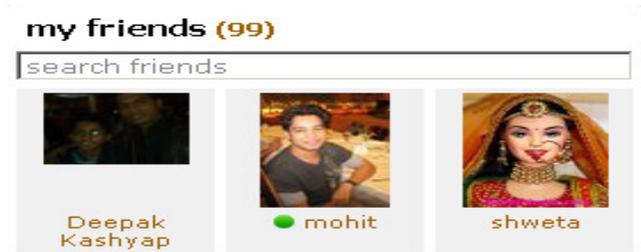

Fig. 1 Friend list

*2) Communities:* - Community is a group of user profiles share common interest. Anyone with an Orkut account can create a community on anything.

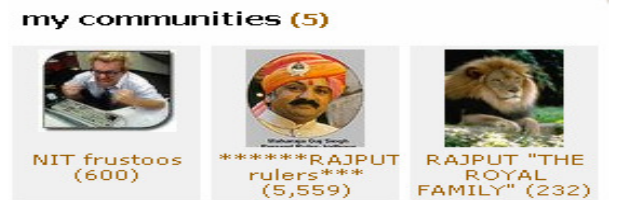

## III. APPROACHES FOR USER PROFILE INVESTIGATION

*A) Identifying connection*

How two profiles are connected and how much closeness both has, this approaches is used to detect the connection between two or more suspicious profiles means how two criminals are connected in web environments.

Users in Web social network visualize as a node and link between users reflect relationship. A user has connection with their friends by maintaining friend list consist of links of all friends profiles.

Our initial approach to link identification consisted of dividing friend's network features into graph features [3].

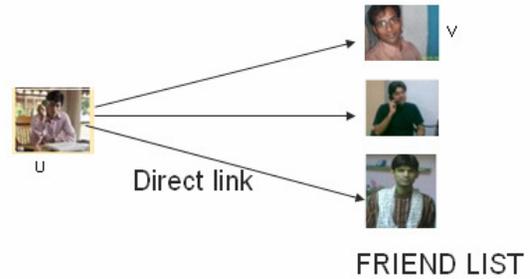

Figure 3 Friends/neighbours

1. in degree of *u*: popularity of the user
2. in degree of *v*: popularity of the candidate
3. Out degree of *u***:** number of other friends besides the Candidate; saturation of friends list
4. Out degree of *v*: number of existing friends of the Candidate besides the user; Correlates loosely with
   Likelihood of a reciprocal link
5. Number of mutual friends w such that $u \rightarrow w \quad w \rightarrow v$
6. "Forward deleted distance": minimum alternative
Distance from *u* to *v* in the graph without the edge (*u*, *v*)
7. Backward distance from *v* to *u* in the graph
These were supplemented by interest-based features:
8. Number of mutual interests between *u* and *v*
9. Number of interests listed by *u*
10. Number of interests listed by *v*
11. Ratio of the number of mutual interests to the number Listed by *u*
12. Ratio of the number of mutual interests to the number Listed by
13-path length: number of links (edges) between u and v in a given path.
14 hop count: number of vertices (users) between u and v in a given path.

*1) Investigate user profile*

findout the strong connection to other profiles Extract friends profiles belongs to same cities, education ,college etc.

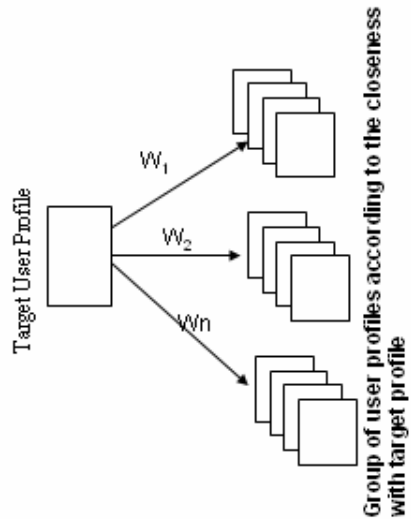

2) *Investigate the relationship link between two or more profiles (used to find-out connection between suspicious profiles)*

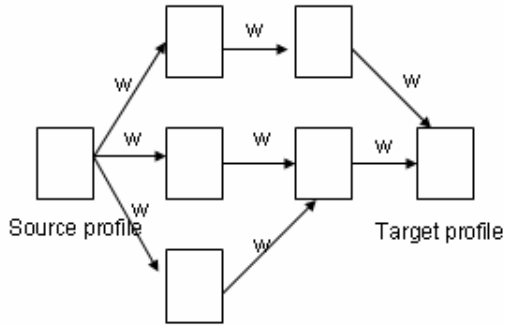

W is the closeness/interaction level

*B) Closeness identification*

W can be calculated by one of the two methods
1) On the basis of communication
2) On the basis of profile similarity.

*1) O n the basis of communication*
In a social network based upon online communication, the distance between individuals does not mean `geographical distance' because each person lives in a virtual world. Instead, distance can be considered `psychological distance' and this can be measured by the influence" wielded among the members of the network. Consider the situation where an individual p has a great deal of influence on an individual q[2].

In this case, we can consider three types of relationship.
Case1: p is close to q.
Case2: q is close to p.
Case3: p and q are close to each other.

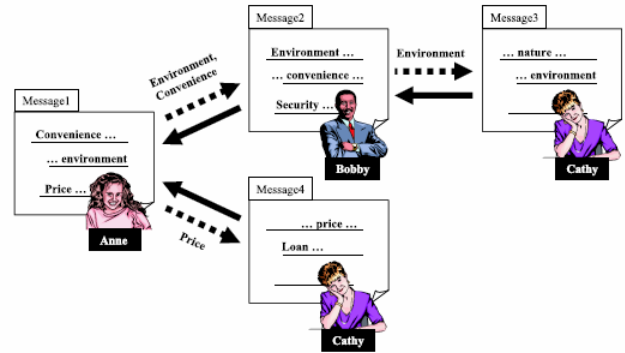

Figure 4: A message chain of messages sent by three individuals.

*2) On the basis of profile similarity.*

2.1) On the basis of contents uniqueness
2.2) On the basis of contents similarity

*2.1) on the basis of contents uniqueness*
Similarity is measured by analyzing text, links. If we are trying to evaluate the likelihood that user A is linked to user B, we sum the number of items the two users have in common. Items that are unique to a few users are weighted more than commonly occurring items. The weighting scheme we use is the inverse log frequency of their occurrence. For example, if only two people mention an item, then the weight of that item is 1/log(2) or 1.4.[1]

$$similarity(A,B) = \sum_{shareditems} \frac{1}{\log[frequency(shareditem)]}$$

It is possible with this algorithm to evaluate each shared item type independently (i.e. links, mailing lists, text) or to combine them together into a single likeness score.

*2.2) on the basis of contents similarity*

Propose work focus on contents based similarity.

### IV. PROPOSED WORK: SIMILARITY MEASUREMENT BASED ON CONTENTS SIMILARITY

Similarity between profiles reflects closeness and interaction between users. Similarity is measured by profile information given by user. Similarity measurement process consists of following.

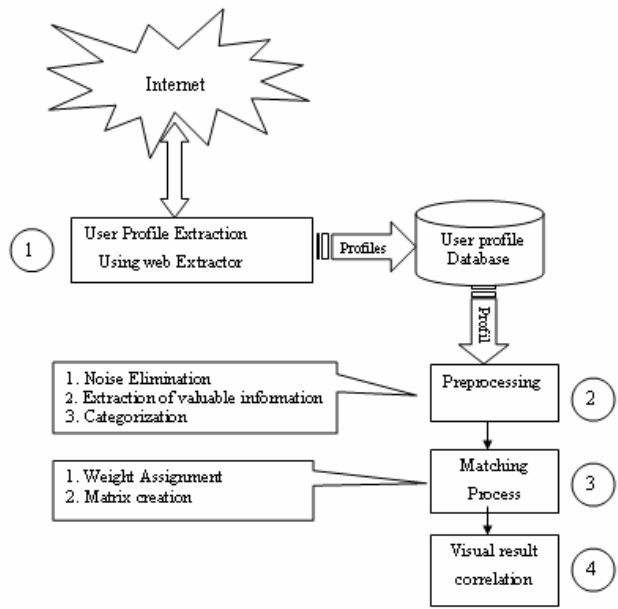

Fig. 5.proposed architecture

http://www.orkut.co.in/Main#FriendsList.aspx?uid=12760208310579966367
http://www.orkut.co.in/Main#FriendsList.aspx?uid=13317425663991525398
http://www.orkut.co.in/Main#FriendsList.aspx?uid=13973160759338911611
http://www.orkut.co.in/Main#FriendsList.aspx?uid=14636742881815046201
http://www.orkut.co.in/Main#FriendsList.aspx?uid=17422035763385012687
http://www.orkut.co.in/Main#Community.aspx?cmm=18034370,
http://www.orkut.co.in/Main#Community.aspx?cmm=8312468

Friends profile ids:-                          Community ids-
12760208310579966367.                    18034370.
13317425663991525398.                    8312468.
13973160759338911611.
14636742881815046201
17422035763385012687

A web data extractor is incorporate to extract profiles from orkut.com. It extracts text and links of a given orkut profile. Text represents profile general information and links represent social connections, like- friends and communities. Web data extractor extract web page that are currently appear on the browser. To extract friends (neighbours) profile we have to browse every individual profile on the browser (internet explorer) and should run the web extractor program manually.

### A. *Extraction of Orkut network.*

Online social networks are part of the Web, but their data representations are very different from general web pages. The Web pages that describe an individual in an online social network are typically well structured, as they are usually automatically generated, unlike general web pages, which could be authored by any person. Therefore, we can be very certain about what pieces of data we can obtain after crawling a particular individual's web pages. In Orkut, links are undirected and link creation requires consent from the target. Since, at the time of the crawl, new users had to be invited by an existing user to join the system. Because Orkut does not export an API, we can resort to HTML screen scraping to conduct crawl Furthermore, Orkut limits the rate at which a single IP address can download information and requires a logged-in account to browse the network. As a result, it took more than a month to crawl million users [5].

From user profile we obtain the user's social, professional, personal information like-religion, ethnicity, age, hometown, city, country, language speak, education, college university ……total 68 field of user profile. Besides this local information, there are usually links that we can use to trace the user's connection to the others, which are hyperlinked to those friends profile pages. Thus, by extracting these hyperlinks, we can construct the graph of connections between all the users in the social network. Every user profile and communities in Orkut social network assigned a unique ID by which they uniquely identified.
Examples: -
Hyperlinks to friend's profiles:-

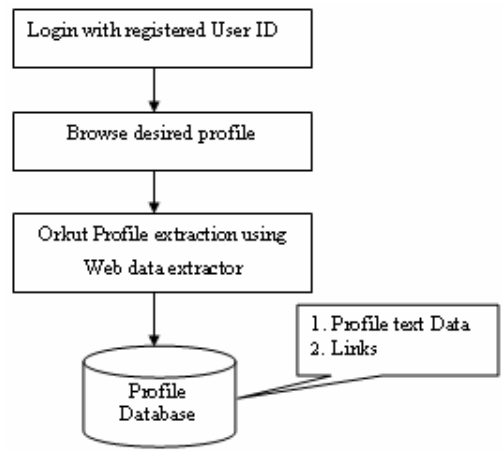

Figure 6 Orkut User profile extraction

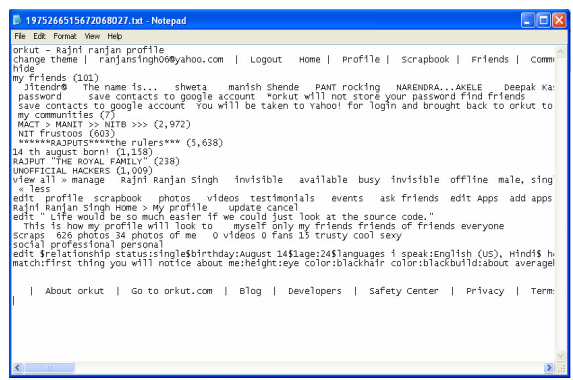

Figure 7 Extracted text file, contains profile Information

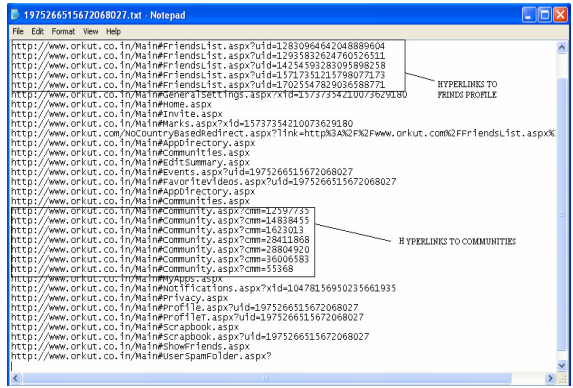

Figure 8 Extracted Link Information (friends profile links, communities' limks)

*B. Pre-Processing.*

Extracted data is in unstructured form. It can not directly used for relationship identification, so data must process and convert into structured form to do this. During resource extraction separate files are created for every orkut profiles contains text and links information. C code has been developed based on token searching approach that takes extracted file as a input and generate output consist of-
1) Friends profile ID's,
2) Communities ID's
3) Users general information like social, professional, educational, interest, Contact.

*1) Extraction of valuable information:* All information of user homepage is not required for similarity measurement. So, only feasible information is extracted. Out of 68 fields 20 fields are considered for similarity measurement

TABLE 3-EXTRACTED PROFILE INFORMATION

| Social | Professional |
|---|---|
| Gender, Relationship status, languages speak, Ethnicity, Religion, Smoking, Drinking, Sports, Hometown, Activity, City, Zip/postal code, State, Country. | Education, College/University, Degree, Occupation, Industry, Company |

*2) Categorization*

Extracted information is classified according to their characteristics.

| Educational and professional Info. | Personal Info. | Interest | Contact Info. |
|---|---|---|---|
| Education | Gender | Sports | Home Town |
| Degree | Language | Activities | PIN Postal Code |
| College/University | Religion | | City |
| Industry | Ethnicity | Smoking | State |
| Occupation | Relationship Status | Drinking | Country |
| Company/organization | | | |

Figure 5 Categorised Information

*C. Matching process.*

Categorised fields are used for similarity measurement. In this proposed work, similarity is measure between a user profile and his friends that are directly linked/ hyperlinked from his homepage. So we calculate similarity weight between a source profiles and his friend's network.

*1) Weight Assignment:*

Some profile fields are better predictor of connection than others, like- if two people mention same city or working in same organization then it shows strong connection as compared to those peoples who mentioned different cities. Professional information like if two people studying in same university/college then college/university field weight would be always higher than education and degree field. So for better prediction of relationship, different weight is assigned to different profile fields. Two methods are proposed to decide weight of different fields:

- Binary weight Assignment
- Weight on the basis of hierarchy

*1.1) Binary Weight Assignment:* Binary weight assigned to every fields.

| Professional & Edu. | Abbreviation | Weight |
|---|---|---|
| Education | $W_{ed}$ | 1/0 |
| Degree | $W_d$ | 1/0 |
| College/University | $W_{cu}$ | 1/0 |
| Industry | $W_{in}$ | 1/0 |
| Occupation | $W_{cv}$ | 1/0 |
| Company | $W_{oc}$ | 1/0 |

| Contact Info. | Abbreviation | Weight |
|---|---|---|
| Home Town | $W_t$ | 1/0 |
| PIN Postal Code | $W_p$ | 1/0 |
| City | $W_c$ | 1/0 |
| State | $W_s$ | 1/0 |
| Country | $W_{co}$ | 1/0 |

| Personal Info. | Abbreviation | Weight |
|---|---|---|
| Gender | $W_g$ | 1/0 |
| language | $W_{la}$ | 1/0 |
| Religion | $W_{re}$ | 1/0 |
| Ethnicity | $W_{et}$ | 1/0 |
| Relationship status | $W_{rs}$ | 1/0 |

| Interest | Abbreviation | Weight |
|---|---|---|
| Sports | $W_{sp}$ | 1/0 |
| Activities | $W_{sm}$ | 1/0 |
| Smoking | $W_{dr}$ | 1/0 |
| Drinking | $W_{ac}$ | 1/0 |

Figure 6 Binary Weight Assignments

Binary weight (0 or 1) is assigned to every fields and these fields weight are multiplied with matching result Discuss in next section. Binary weight can be used to mask the result (to show something or to hide something).

*1.2) Hierarchy based weight assignment.*

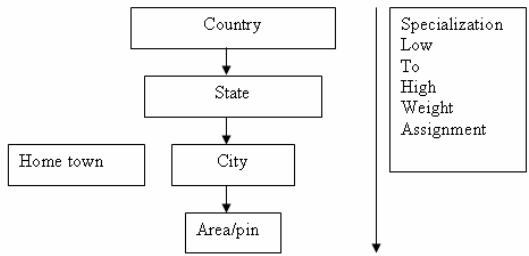

Figure 9: Contact Information Hierarchy

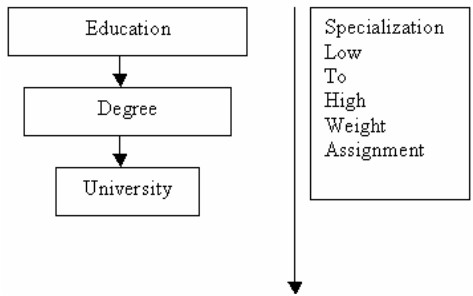

Figure 10: Educational information Hierarchy

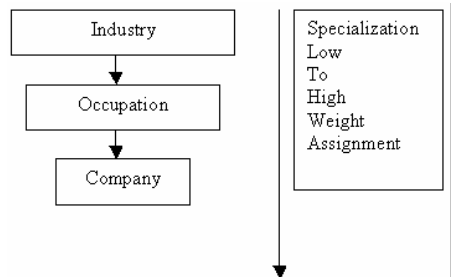

Figure 11: Professional information Hierarchy.

Weight is assigned according to the position in hierarchy

| Contact Info. | Weight |
|---|---|
| Home Town | 3 |
| PIN Postal Code | 4 |
| City | 3 |
| State | 2 |
| Country | 1 |

| Job description | Weight |
|---|---|
| Industry | 1 |
| Occupation | 2 |
| Company/organization | 3 |

| Educational Info. | Weight |
|---|---|
| Education | 1 |
| Degree | 2 |
| College/University | 3 |

Figure 12 Hierarchy Based Weight Assignment

Some categories doesn't form hierarchy like interest, personal. Whereas some categories like contact, professional and educational forms hierarchy. So that, the weight is assigned according to the position in hierarchy.

*2). Matching Matrix/ Table Creation.*

Separate mating matrix has created for every category of information.

TABLE 4. GENERAL MATRIX FORMAT

| Friends | Field1 | Field2 | Field N | Total Weight |
|---|---|---|---|---|
| f1 | 0/1 | 0/1 | 0/1 | |
| f2 | 0/1 | 0/1 | 0/1 | |
| . | . | . | . | |
| Fn | 0/1 | 0/1 | 0/1 | |

0-Match, 1-No Match

Table 6 shows the general format of matching matrix. Columns specify the fields and rows specify the friends matching result, in the form of 0/1 (1-for perfect Match, 0-No match).

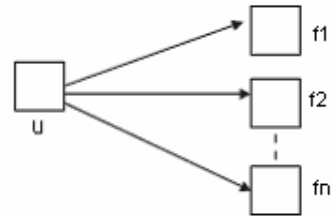

Figure 13 User connections

f1, f2,…,fn are friends of U.U is a source profile from which we perform matching.

TABLE 5: CONTACT INFO MATCHING MATRIX/TABLE.

| Friends | Home town | Pin code | City | State | Country | Total Weight |
|---|---|---|---|---|---|---|
| f1 | 0/1 | 0/1 | 0/1 | 0/1 | 0/1 | $W_{c1}$ |
| f2 | 0/1 | 0/1 | 0/1 | 0/1 | 0/1 | $W_{c2}$ |
| . | . | . | . | . | . | . |
| fn | 0/1 | 0/1 | 0/1 | 0/1 | 0/1 | $W_{cn}$ |
| | X $W_t$ | X $W_p$ | X $W_c$ | X $W_s$ | X Wco | |

$W_t$, $W_p$, $W_c$, $W_s$, $W_{co}$ are the weight of hometown, pin code, city, state, country respectively and these weight will be multiplied with the matching result. $W_{c1}$, $W_{c2},...,W_{cn}$ are total matching score. Matching matrix specifies the similarity between U and their friends according to the similarity in contact information.

TABLE 6: EDUCATIONAL & PROFESSIONAL INFO. MATCHING MATRIX/TABLE

| Friends | Education | Degree | College | Industry |
|---|---|---|---|---|
| $F_1$ | 0/1 | 0/1 | 0/1 | 0/1 |
| $F_2$ | 0/1 | 0/1 | 0/1 | 0/1 |
| . | . | . | . | . |
| $F_n$ | 0/1 | 0/1 | 0/1 | 0/1 |
|  | X | X | X | X |
|  | $W_{ed}$ | $W_d$ | $W_{cu}$ | $W_{in}$ |

| Occupation | Company | Total Weight |
|---|---|---|
| 0/1 | 0/1 | $Wp_1$ |
| 0/1 | 0/1 | $Wp_2$ |
| . | . | . |
| 0/1 | 0/1 | $Wp_n$ |
| X | X |  |
| $W_{cy}$ | $W_{oc}$ |  |

$W_{ed}$, $W_d$, $W_{cu}$, $W_{in}$, $W_{cy}$, $W_{oc}$, are the weight of fields Education, degree, college/university, industry, occupation company, respectively and will be multiplied with the matching result (1/0). $W_{p1}$, $W_{p2}$,…,$W_{pn}$ are total matching score. Matching matrix specify the similarity between U and their friends according to the similarity in professional & Educational information.

TABLE17: PERSONAL INFORMATION matching matrix/TABLE

| Fri. | Gender | Language | Ethnicity | Religion | Rela. Status | Tot. Weight |
|---|---|---|---|---|---|---|
| f1 | 0/1 | 0/1 | 0/1 | 0/1 | 0/1 | $W_{pe1}$ |
| f2 | 0/1 | 0/1 | 0/1 | 0/1 | 0/1 | $W_{pe2}$ |
| . | . | . | . | . | . | . |
| Fn | 0/1 | 0/1 | 0/1 | 0/1 | 0/1 | $W_{pen}$ |
|  | X | X | X | X | X |  |
|  | $W_g$ | $W_{la}$ | $W_{et}$ | $W_{re}$ | $W_{rs}$ |  |

$W_g$, $W_{la}$, $W_{et}$, $W_{re}$, $W_{rs}$ are the weight of fields Gender, Language, Ethnicity, Religion, Relationship status respectively and will be multiplied with the matching result (1/0). $W_{pe1}$, $W_{pe2}$, $W_{pen}$ are total matching score of friend f1,f2,fn respectively with U. matching matrix specify the similarity between U and their friends according to the similarity in Personal information.

TABLE8: INTEREST INFORMATION matching matrix

| Fri. | Sports | Smoking | Drinking | Activities | Total Wei. |
|---|---|---|---|---|---|
| f1 | 0/1 | 0/1 | 0/1 | 0/1 | $W_{i1}$ |
| f2 | 0/1 | 0/1 | 0/1 | 0/1 | $W_{i2}$ |
| . | . | . | . |  | . |
| fn | 0/1 | 0/1 | 0/1 | 0/1 | $W_{in}$ |
|  | X | X | X | X |  |
|  | $W_{sp}$ | $W_{sm}$ | $W_{dr}$ | $W_{ac}$ |  |

$W_{sp}$, $W_{sm}$, $W_{dr}$, $W_{ac}$ are the weight of fields Sports, Smoking, Drinking,, Activities respectively and will be multiplied with the matching result (1/0). $W_{i1}$, $W_{i2}$,$W_{in}$ are total matching score of friend f1,f2,fn respectively with U. matching matrix specify the similarity between U and their friends according to the similarity in user interest information.

*3) Weight of mutual friends & mutual communities*
*1) Mutual friends:* - Specify the mutual social connection between users. When two friends having greater number of mutual friends then they create mutual social networks. So for finding closeness, weights of mutual friends are added in resultant similarity score.
Number of mutual friends w such that $u \rightarrow w \quad w \rightarrow fi$

$$Mfi = \frac{\frac{Mutual\ friends(U,fi)}{Total\ No.\ of\ U's\ friends} * 100}{WAF}$$

U – Base profile, and $f_i$- indicate friends profiles.
$Mf_i$- mutual friends weight between u and $f_i$.
WAF- weight adjustment factor decides the upper limit of the weight, if WAF is 10, then maximum weight of mutual friends cannot exceed above 10.

*2) Mutual communities:* -specify the mutual interest between users. Users may belong to any number of communities that reflect his interest. When two friends have greater number of mutual communities then they are more closed according to shared interest. So, weights of mutual communities are added into resultant weight.

$$Mci = \frac{\frac{Mutual\ communities(U,fi)}{Total\ No.\ of\ U's\ Communities} * 100}{WAF}$$

U – Base profile.
$f_i$- indicates friends profiles.
$Mc_i$- mutual communities weight of user i.
Mutual communities (u, $f_i$) = mutual communities between u and fi
WAF- weight adjustment factor defines the upper limit of weight, if WAF is 10 then maximum weight of mutual communities can not exceed above 10.

*D. Visual result Correlation:*

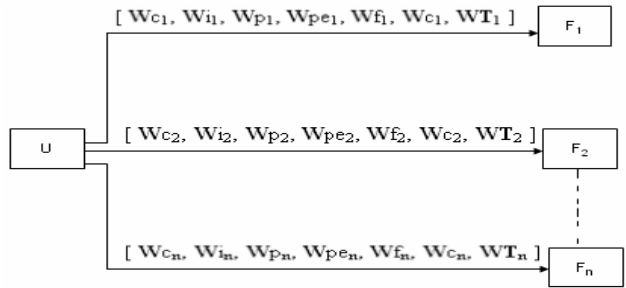

Fig.14. Resultant Similarity Weight

$$\sum_{i=0}^{n} WT_i = W_{ci} + W_{ii} + W_{pi} + W_{pei} + W_{fi} + W_{ci} + WT_i$$

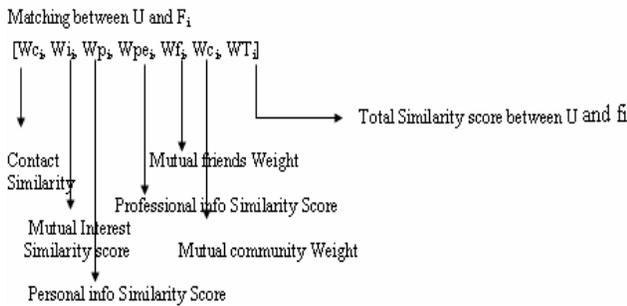

Wc1, Wc2…. Wcn-Contact information similarity score.
Wp1, Wp2… Wpn-Professional Similarity matching Score.
Wpe1, Wpe2..Wpen-Personal Info. Similarity Matching Score.
Wi1, Wi2……Win-Interest Matching Score
Mf1, Mf2…….Mfn-Mutual friends Weight.
Mcu1,Mcu2...Mcu-Mutual Communities weight.
TW1, TW2, TWn=total similarity weight between U and f1,f2,fn restrictively.
$WT_1 = W_{c1} + W_{i1} + W_{p1} + W_{pe1} + W_{f1} + W_{c1} + WT1$
$WT_2 = W_{c2} + W_{i2} + W_{p2} + W_{pe2} + W_{f2} + W_{c2} + WT2$
$WT_n = W_{cn} + W_{in} + W_{pn} + W_{pen} + W_{fn} + W_{cn} + WT_n$

## V. EXPERIMENT RESULTS

Data is gathered from www.orkut.com. In this work 9 Orkut profiles are used as sample data, extracted by using web data extractor. Pre-processing and matching has been performed by pre-processing engine developed in C language (using file handing & string function). Token searching approach is used to extract useful patterns. In this experiment, matching is performed between a base profiles (ex. Rajni ranjan singh) and their neighbours profiles. One-to-many profile matching is performed.

```
==========================================
Name=Rajni Ranjan Singh profile
Male
userid=19752665156720 68027
==========================================
Name= Neetu's profile
Female
Userid=15772385635974559064
==========================================
Name= akhtar's profile
Male
userid=18172151621177275498
==========================================
Name= DEVVRIT's profile
Male
Userid=1138302238349976680
==========================================
Name= Priyanka's profile
Female
Userid=12898686644814692716
==========================================
Name= NikhilPaliCISCOa's profile
Userid=913409904372138593
Male
==========================================
Name= NILESH PATEL's profile
Male
Userid=548768002074853061
==========================================
Name= Abhilash's profile
Male
Userid=17025547829036588771
==========================================
Name= rahul~csc/MANIT's profile
Male
Userid=14696586233830216050
==========================================
Name= Gourishankar's profile
Male
Userid=10002147457212214349
```

**Binary Weight Assignment**

Contact info. Matching Matrix.

| UID | hometown | Area code | city | state | country | total |
|---|---|---|---|---|---|---|
| 15772385635974559064 | 0 | 0 | 1 | 1 | 1 | 3 |
| 18172151621177275498 | 0 | 0 | 0 | 0 | 1 | 1 |
| 1138302238349976680 | 1 | 1 | 1 | 1 | 1 | 5 |
| 12898686644814692716 | 0 | 0 | 1 | 1 | 1 | 3 |
| 913409904372138593 | 0 | 0 | 0 | 0 | 1 | 1 |
| 548768002074853061 | 0 | 0 | 1 | 1 | 1 | 3 |
| 17025547829036588771 | 0 | 0 | 0 | 0 | 1 | 1 |
| 14696586233830216050 | 0 | 0 | 1 | 1 | 1 | 3 |
| 10002147457212214349 | 0 | 0 | 0 | 1 | 1 | 2 |

| UID | Lan.Speak | religion | Ethnicity | sex | status | total |
|---|---|---|---|---|---|---|
| 15772385635974559064 | 0 | 1 | 0 | 0 | 0 | 1 |
| 18172151621177275498 | 1 | 0 | 0 | 1 | 1 | 3 |
| 1138302238349976680 | 1 | 1 | 0 | 1 | 1 | 4 |
| 12898686644814692716 | 0 | 0 | 0 | 0 | 1 | 1 |
| 913409904372138593 | 1 | 1 | 1 | 1 | 1 | 5 |
| 548768002074853061 | 1 | 1 | 0 | 1 | 0 | 3 |
| 17025547829036588771 | 1 | 1 | 1 | 1 | 1 | 5 |
| 14696586233830216050 | 0 | 1 | 0 | 1 | 0 | 2 |
| 10002147457212214349 | 1 | 1 | 0 | 1 | 1 | 4 |

| UID | smoking | drinking | sports | activities | total |
|---|---|---|---|---|---|
| 15772385635974559064 | 0 | 0 | 0 | 0 | 0 |
| 18172151621177275498 | 0 | 0 | 0 | 0 | 0 |
| 11383022383499766680 | 0 | 0 | 1 | 0 | 1 |
| 12898686644814692716 | 0 | 0 | 1 | 0 | 1 |
| 913409904372138593 | 0 | 0 | 1 | 0 | 1 |
| 548768002074853061 | 0 | 0 | 1 | 0 | 1 |
| 17025547829036588771 | 0 | 0 | 0 | 0 | 0 |
| 14696586233830216050 | 0 | 0 | 0 | 0 | 0 |
| 10002147457212214349 | 0 | 0 | 1 | 0 | 1 |

| UID | education | degree | college | industry | company | occupation | total |
|---|---|---|---|---|---|---|---|
| 15772385635974559064 | 0 | 0 | 0 | 0 | 0 | 0 | 0 |
| 18172151621177275498 | 0 | 0 | 0 | 0 | 0 | 0 | 0 |
| 11383022383499766680 | 0 | 0 | 0 | 0 | 0 | 0 | 0 |
| 12898686644814692716 | 0 | 0 | 0 | 0 | 0 | 1 | 1 |
| 913409904372138593 | 1 | 0 | 0 | 0 | 0 | 0 | 1 |
| 548768002074853061 | 1 | 0 | 0 | 1 | 0 | 0 | 2 |
| 17025547829036588771 | 1 | 0 | 0 | 1 | 0 | 1 | 3 |
| 14696586233830216050 | 1 | 0 | 0 | 0 | 0 | 0 | 1 |
| 10002147457212214349 | 0 | 0 | 0 | 0 | 0 | 0 | 0 |

| UID | mutual friend | mutual communities |
|---|---|---|
| 15772385635974559064 | 1 | 2 |
| 18172151621177275498 | 1 | 0 |
| 11383022383499766680 | 3 | 0 |
| 12898686644814692716 | 1 | 0 |
| 913409904372138593 | 2 | 0 |
| 548768002074853061 | 2 | 3 |
| 17025547829036588771 | 2 | 2 |
| 14696586233830216050 | 2 | 2 |
| 10002147457212214349 | 2 | 2 |

| UID | Contact | Personal | Interest | Education | Mu.Fri | Mu.comm | Total |
|---|---|---|---|---|---|---|---|
| 15772385635974559064 | 3 | 1 | 0 | 0 | 1 | 2 | 7 |
| 18172151621177275498 | 1 | 3 | 0 | 0 | 1 | 0 | 5 |
| 11383022383499766680 | 5 | 4 | 1 | 0 | 3 | 0 | 13 |
| 12898686644814692716 | 3 | 1 | 1 | 1 | 1 | 0 | 7 |
| 913409904372138593 | 1 | 5 | 1 | 1 | 2 | 0 | 10 |
| 548768002074853061 | 3 | 3 | 1 | 2 | 2 | 3 | 14 |
| 17025547829036588771 | 1 | 5 | 0 | 3 | 2 | 2 | 13 |
| 14696586233830216050 | 3 | 2 | 0 | 1 | 2 | 2 | 10 |
| 10002147457212214349 | 2 | 4 | 1 | 0 | 2 | 2 | 11 |

**On the basis of hierarchy**

| UID | Contact | Personal | Interest | Education | Mu.Fri | Mu.comm | Total |
|---|---|---|---|---|---|---|---|
| 15772385635974559064 | 6 | 1 | 0 | 0 | 1 | 2 | 10 |
| 18172151621177275498 | 1 | 3 | 0 | 0 | 1 | 0 | 5 |
| 11383022383499766680 | 13 | 4 | 1 | 0 | 3 | 0 | 21 |
| 12898686644814692716 | 6 | 1 | 1 | 2 | 1 | 0 | 11 |
| 913409904372138593 | 1 | 5 | 1 | 1 | 2 | 0 | 10 |
| 548768002074853061 | 6 | 3 | 1 | 2 | 2 | 3 | 17 |
| 17025547829036588771 | 1 | 5 | 0 | 4 | 2 | 2 | 14 |
| 14696586233830216050 | 6 | 2 | 0 | 1 | 2 | 2 | 13 |
| 10002147457212214349 | 3 | 4 | 1 | 0 | 2 | 2 | 12 |

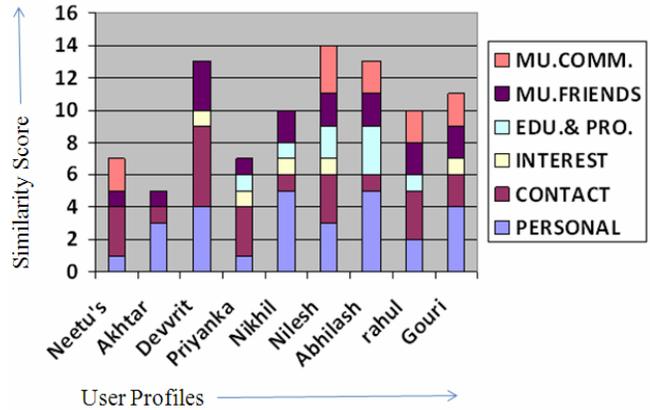

Chart I: Similarity Score using binary weight assignment

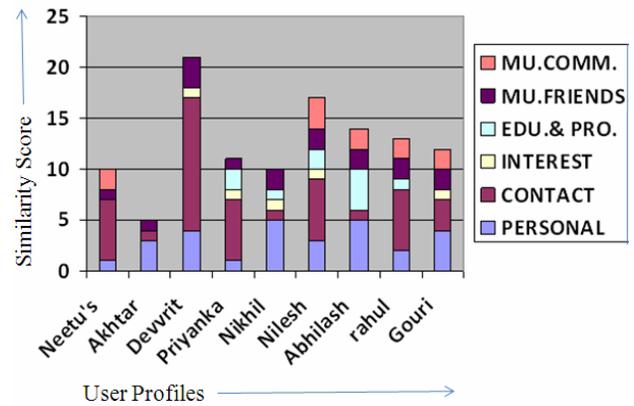

Chart II: Similarity Score using Hierarchical weight assignment

Two result Matrix has been generated. One is by using binary weight and another one is by using hierarchical weight assignment. Total similarity score is the successive addition of each categorized matching results. As per shown in chart 1, *Nilesh* profile is closer to the base profile (*Rajni ranjan* profile) since personal, contact and mutual communities' fields are more similar to base profile. In chart II, *Devvrit* profile is closer to base profile in which contact information play an important role.

## VI. CHALLENGES

*A) Orkut privacy issue and extraction of meaningful patterns*

Of course, not all users provide their social, professional, personal, interest's information, and even if they did, privacy settings may prevent us from viewing their profile contents. This availability of data was not a *huge* problem, but it could potentially skew our ability to extract meaningful patterns. The bigger issue, however, is the natural language processing problem: we are ultimately interested in the *meaning* behind the words. If different users list "Madhya Pradesh" and "M.P", same as in College/University field users list MANIT and Maulana azad NIT" and "NITB" it is hard for us to realize that these are the same state and same college. Plus, this is a syntactical issue; more interesting situations are when different users may list interest (Activities): "jogging through the streets,""jogging,""jogging!," and" jogging, but only on a treadmill." Even as a human it can be difficult to determine how specific we should be when classifying. Last, even if we only look for keywords, there could be errors: "school," "anything but school" is clearly not the same. To reduce these problem we try to work on data that are come from fields contains only predefined set of data, in programming term we can say data taken from Combo boxes like country, religion, ethnicity, smoking, Drinking, Gender, Relationship Status ,Education, Degree, Industry ,language speak.

Web social networks are dynamic by nature user may add more friends and join many communities and can change profiles contents so similarity score may change according to time.

## VII CONCLUSION

*This paper aims to answer the question: Are social links valid indicators of real user interaction?* Profile based similarity show the exact relationship between users. Similarity between two-user profiles on Orkut is measured on the basis of social, geographical, educational, professional, shared interest (including mutual communities) and mutual social connection (mutual friends). The measured similarity score may be used as a trust between users. Profile based similarity measurement is useful for investigation of users profile. Similarity between user profiles reflects closeness and interaction between users.

## VII REFERENCES

[1] Adamic L.A. and Adar. E "Friends and Neighbors on the web" Social Networks,Vol 25,2007,pp 211-230.

[2] Mining Directed Social Network from Message Board, Naohiro Matsumura Osaka University,David E. Goldberg UIUC,Xavier Llor `aUIUC WWW 2005, May 10–14, 2005, Chiba, Japan.ACM 1595930515/05/0005.

[3] Structural Link Analysis from User Profiles and Friends Networks: A Feature Construction Approach William H. Hsu Joseph Lancaster Martin S.R. Paradesi Tim Weninger ICWSM'2007 Boulder, Colorado, USA.

[4] Discover behaviour of Turkish people in Orkut Alberto Ochoa-Zezzatti, Javier Martínez, Alberto Hernández & Jaime Muñoz,UAIE UAZ / Instituto Tecnológico de León (Maestría en Sistemas Inteligentes).17[th] international conference on electronics ,communication and computers(CONIELE COMP'07)) 0-7695-2799-x/07 2007 IEEE

[5] Measurement and Analysis of Online Social Networks ,Alan Mislove,Massimiliano Marcon, Krishna P. Gummadi, Peter Druschel, Bobby Bhattacharjee, IMC'07, October 24-26, 2007, San Diego, California, USA. Copyright 2007 ACM 978-1-59593-908-1/07/0010.

[6] "An Integrated method for social network extraction" Masahiro Hamasak, Yutaka Matsuo,WWW 2006 may 23-26,2006 ACM 1-59593-332-9/06/0005.

[7] Trust and Nuanced Profile Similarity in Online Social Networks Jennifer Golbeck_ golbeck@cs.umd.edu University of Maryland, College Park 8400 Baltimore Avenue, Suite 200, College Park, Maryland 20740

[8]A Framework to identify relationship among user profile in Web Social Environment,Deepak singh Tomar,Sc shrivastava,Rajni Ranjan Singh, 2[nd] National Conference on Emerging Principal and Practices of Computer Science and Information Technology (EPPCSIT 2009) at Ludianna:ISBN:81-89652-36-6.

**Authors profile:**

**Mr. Rajni Ranjan Singh**, received M Tech in Information Security from Maulana Azad National Institute of Technology Bhopal India in 2009. He obtained his BE in Computer Science and Engineering from SATI Vidisha in 2006. His research activities based on, System and network security, Web Technology.

**Mr. Deepak Singh Tomar**, working as Assistant Professor in Department of Computer Science at MANIT Bhopal. His research activities are based on digital forensics, data mining and network security.